%
%
\magnification=\magstep0
\headline={\ifnum\pageno=1\hfil\else\hfil\tenrm--\ \folio\ --\hfil\fi}
\footline={\hfil}
\hsize=6.0truein
\vsize=8.54truein
\hoffset=0.25truein
\voffset=0.25truein
\baselineskip=15pt
%
%
\tolerance=9000
\hyphenpenalty=10000
%
%
%

\font\mbf=cmmib10 \font\mbfs=cmmib10 scaled 833
\font\msybf=cmbsy10 \font\msybfs=cmbsy10 scaled 833

%
%
%

\textfont9=\mbf \scriptfont9=\mbfs \scriptscriptfont9=\mbfs

\textfont10=\msybf \scriptfont10=\msybfs \scriptscriptfont10=\msybfs
%
%
\mathchardef\alpha="710B
\mathchardef\beta="710C
\mathchardef\gamma="710D
\mathchardef\delta="710E
\mathchardef\epsilon="710F
\mathchardef\zeta="7110
\mathchardef\eta="7111
\mathchardef\theta="7112
\mathchardef\iota="7113
\mathchardef\kappa="7114
\mathchardef\lambda="7115
\mathchardef\mu="7116
\mathchardef\nu="7117
\mathchardef\xi="7118
\mathchardef\pi="7119
\mathchardef\rho="711A
\mathchardef\sigma="711B
\mathchardef\tau="711C
\mathchardef\upsilon="711D
\mathchardef\phi="711E
\mathchardef\chi="711F
\mathchardef\psi="7120
\mathchardef\omega="7121
\mathchardef\varepsilon="7122
\mathchardef\vartheta="7123
\mathchardef\varpi="7124
\mathchardef\varrho="7125
\mathchardef\varsigma="7126
\mathchardef\varphi="7127
\mathchardef\nabla="7272
\mathchardef\cdot="7201
%
%
\def\spose#1{\hbox to 0pt{#1\hss}}
\def\lta{\mathrel{\spose{\lower 3pt\hbox{$\mathchar"218$}}
     \raise 2.0pt\hbox{$\mathchar"13C$}}}
\def\gta{\mathrel{\spose{\lower 3pt\hbox{$\mathchar"218$}}
     \raise 2.0pt\hbox{$\mathchar"13E$}}}
%
%

%
%

%
%

%
%
%

\def\cm{{\rm\,cm}}

\def\pc{{\rm\,pc}}
\def\kpc{{\rm\,kpc}}

\def\yr{{\rm\,yr}}

\def\kms{{\rm\,km\,s^{-1}}}

\def\msun{{\,M_\odot}}
\def\lsun{{\,L_\odot}}

\def\gev{{\rm\,GeV}}

\def\etal{et al.$~$}

\def\muas{\, \mu{\rm as}}
\def\mas{{\, \rm mas}}
%
%
\def\degree{^\circ}

%
%
\newcount\eqnumber
\eqnumber=1
%
\def\new{{\the\eqnumber}\global\advance\eqnumber by 1}
%
%
\def\ref#1{\advance\eqnumber by -#1 \the\eqnumber
     \advance\eqnumber by #1 }
%
%
\def\last{\advance\eqnumber by -1 {\the\eqnumber}\advance 
     \eqnumber by 1}
%
%
\def\eqnam#1{\xdef#1{\the\eqnumber}}
%
%
%
\def\refindent{\par\noindent\hangindent=3pc\hangafter=1 }
\def\aa#1#2#3{\refindent#1, A\&A, #2, #3}

\def\actastr#1#2#3{\refindent#1, Acta Astr., #2, #3}
\def\aj#1#2#3{\refindent#1, AJ, #2, #3}

\def\apj#1#2#3{\refindent#1, ApJ, #2, #3}

\def\araa#1#2#3{\refindent#1, ARA\&A, #2, #3}

\def\nature#1#2#3{\refindent#1, Nature, #2, #3}

\def\physrevl#1#2#3{\refindent#1, Phys. Rev. Lett., #2, #3}

\def\science#1#2#3{\refindent#1, Science, #2, #3}

\def\refbook#1{\refindent#1}

%
%
\def\refrule{\hbox to 3pc{\leaders\hrule depth-2pt height 2.4pt\hfill}}
%
%
%
\def\sect#1 {
  \vskip 1. truecm plus .2cm
  \bigbreak
  \centerline{\bf #1}
  \nobreak
  \bigskip
  \nobreak}
\def\subsec#1#2 {
  \bigbreak
  \centerline{#1.~{\bf #2}}
  \bigskip}
%
%

%
\newread\epsffilein    
\newif\ifepsffileok    
\newif\ifepsfbbfound   
\newif\ifepsfverbose   
\newdimen\epsfxsize    
\newdimen\epsfysize    
\newdimen\epsftsize    
\newdimen\epsfrsize    
\newdimen\epsftmp      
\newdimen\pspoints     
\pspoints=1bp          
\epsfxsize=0pt         
\epsfysize=0pt         
\def\epsfbox#1{\global\def\epsfllx{72}\global\def\epsflly{72}%
   \global\def\epsfurx{540}\global\def\epsfury{720}%
   \def\lbracket{[}\def\testit{#1}\ifx\testit\lbracket
   \let\next=\epsfgetlitbb\else\let\next=\epsfnormal\fi\next{#1}}%
\def\epsfgetlitbb#1#2 #3 #4 #5]#6{\epsfgrab #2 #3 #4 #5 .\\%
   \epsfsetgraph{#6}}%
\def\epsfnormal#1{\epsfgetbb{#1}\epsfsetgraph{#1}}%
\def\epsfgetbb#1{%
%
%
\openin\epsffilein=#1
\ifeof\epsffilein\errmessage{I couldn't open #1, will ignore it}\else
%
%
   {\epsffileoktrue \chardef\other=12
    \def\do##1{\catcode`##1=\other}\dospecials \catcode`\ =10
    \loop
       \read\epsffilein to \epsffileline
       \ifeof\epsffilein\epsffileokfalse\else
%
%
          \expandafter\epsfaux\epsffileline:. \\%
       \fi
   \ifepsffileok\repeat
   \ifepsfbbfound\else
    \ifepsfverbose\message{No bounding box comment in #1; using defaults}\fi\fi
   }\closein\epsffilein\fi}%
%
%
\def\epsfclipstring{}
\def\epsfsetgraph#1{%
   \epsfrsize=\epsfury\pspoints
   \advance\epsfrsize by-\epsflly\pspoints
   \epsftsize=\epsfurx\pspoints
   \advance\epsftsize by-\epsfllx\pspoints
%
%
   \epsfxsize\epsfsize\epsftsize\epsfrsize
   \ifnum\epsfxsize=0 \ifnum\epsfysize=0
      \epsfxsize=\epsftsize \epsfysize=\epsfrsize
      \epsfrsize=0pt
%
%
     \else\epsftmp=\epsftsize \divide\epsftmp\epsfrsize
       \epsfxsize=\epsfysize \multiply\epsfxsize\epsftmp
       \multiply\epsftmp\epsfrsize \advance\epsftsize-\epsftmp
       \epsftmp=\epsfysize
       \loop \advance\epsftsize\epsftsize \divide\epsftmp 2
       \ifnum\epsftmp>0
          \ifnum\epsftsize<\epsfrsize\else
             \advance\epsftsize-\epsfrsize \advance\epsfxsize\epsftmp \fi
       \repeat
       \epsfrsize=0pt
     \fi
   \else \ifnum\epsfysize=0
     \epsftmp=\epsfrsize \divide\epsftmp\epsftsize
     \epsfysize=\epsfxsize \multiply\epsfysize\epsftmp   
     \multiply\epsftmp\epsftsize \advance\epsfrsize-\epsftmp
     \epsftmp=\epsfxsize
     \loop \advance\epsfrsize\epsfrsize \divide\epsftmp 2
     \ifnum\epsftmp>0
        \ifnum\epsfrsize<\epsftsize\else
           \advance\epsfrsize-\epsftsize \advance\epsfysize\epsftmp \fi
     \repeat
     \epsfrsize=0pt
    \else
     \epsfrsize=\epsfysize
    \fi
   \fi
%
%
   \ifepsfverbose\message{#1: width=\the\epsfxsize, height=\the\epsfysize}\fi
   \epsftmp=10\epsfxsize \divide\epsftmp\pspoints
   \vbox to\epsfysize{\vfil\hbox to\epsfxsize{%
      \ifnum\epsfrsize=0\relax
        \includegraphics{#1}%
      \else
        \epsfrsize=10\epsfysize \divide\epsfrsize\pspoints
        \includegraphics{#1}%
      \fi
      \hfil}}%
\global\epsfxsize=0pt\global\epsfysize=0pt}%
%
%
{\catcode`\%=12 \global\let\epsfpercent=
%
%
\long\def\epsfaux#1#2:#3\\{\ifx#1\epsfpercent
   \def\testit{#2}\ifx\testit\epsfbblit
      \epsfgrab #3 . . . \\%
      \epsffileokfalse
      \global\epsfbbfoundtrue
   \fi\else\ifx#1\par\else\epsffileokfalse\fi\fi}%
%
%
\def\epsfempty{}%
\def\epsfgrab #1 #2 #3 #4 #5\\{%
\global\def\epsfllx{#1}\ifx\epsfllx\epsfempty
      \epsfgrab #2 #3 #4 #5 .\\\else
   \global\def\epsflly{#2}%
   \global\def\epsfurx{#3}\global\def\epsfury{#4}\fi}%
%
%
\def\epsfsize#1#2{\epsfxsize}
%
%

\newcount\notenumber
\notenumber=1
\def\note#1{\footnote{$^{\the\notenumber}$}{#1}\global\advance\notenumber by 1}

\def\etal{{et al.}\ }

\font\gkvec=cmmib10                         

\def\balpha{\hbox{{\gkvec\char11}}}	   
\def\btheta{\hbox{{\gkvec\char18}}}        
\def\bmu{\hbox{{\gkvec\char22}}}	   

\parskip .15cm plus .1cm
\null\vskip 1.cm

\centerline{\bf MICROLENSING EVENTS}
\bigskip
\centerline{\bf FROM MEASUREMENTS OF THE DEFLECTION}
\bigskip
\bigskip
\centerline{Jordi Miralda-Escud\'e\footnote{$^1$}{Institute for Advanced Study,
Princeton, NJ 08540} }
\centerline{E-mail: jordi@ias.edu}
\vskip 0.5in
\centerline{Submitted to The Astrophysical Journal (Letters)}
\centerline{May 17, 1996}

\vskip 1.cm

\sect{ABSTRACT}

  Microlensing events are now regularly being detected by monitoring
the flux of a large number of potential sources and measuring the
combined magnification of the images. This phenomenon could also be
detected directly from the gravitational deflection, by means of high
precision astrometry using interferometry. Relative astrometry at the
level of $10\muas$ may become possible in the near future. The
gravitational deflection can be measured by astrometric monitoring of
a bright star having a background star within a small angular
separation. This type of monitoring program will be carried out for the
independent reasons of discovering planets from the angular motion they
induce on the nearby star around which they are orbiting, and for
measuring parallaxes, proper motions and orbits of binary stars.
We discuss three applications of the measurement of gravitational
deflections by astrometric monitoring: measuring the mass of
the bright stars that are monitored, measuring the mass of brown
dwarfs or giant planets around the bright stars, and detecting
microlensing events by unrelated objects near the line of sight to the
two stars. We discuss the number of stars whose mass could be measured
by this procedure. We also give expressions for the number of expected
microlensing events by unrelated objects, which could be stars, brown
dwarfs, or other compact objects accounting for dark matter in the halo
or in the disk.

\noindent{{\it Subject headings}: gravitational lensing - astrometry -
techniques: interferometric - stars: masses - stars: brown dwarfs -
planetary systems}

\vfill\eject

\sect{1. INTRODUCTION}

  The search for gravitational microlensing in stars of the Large
Magellanic Cloud was suggested by Paczy\'nski (1986) as a technique to
discover compact objects that might account for part of the dark matter.
Microlensing in the galactic bulge can be similarly used to study the
distribution and mass function of stars (or dark matter) in our Galaxy
(Paczy\'nski 1991; Griest 1991; Kiraga \& Paczy\'nski 1994; Zhao,
Spergel, \& Rich 1995; Han \& Gould 1996. See also Paczy\'nski 1996 for
a review). About 100 microlensing events have been detected so far over
three years, mostly towards the bulge (Udalski \etal 1994a,b,c; Alcock
\etal 1995, 1996a,b; Alard 1996). In principle, all the stars in our
Galaxy can be microlensed by other stars in the foreground, although the
optical depth is generally much lower than towards the bulge.

  An alternative technique to monitoring the flux of a large number of
potential sources to detect microlensing events is to search for
candidate lenses, and then check if there are any sources along the path
of the lens once the proper motion is known. This only works for lenses
with high proper motion, in which case the positions of candidate
sources can be measured and used to predict the event a reasonable time
before it takes place. Paczy\'nski (1995) has proposed to search for
such high proper motion stars in the bulge fields, or elsewhere in the
galactic plane; in many cases, these stars will be faint M dwarfs which,
even if they are nearby, will still not be much brighter than many field
stars at the distance of the bulge, and microlensing could be observed
from the usual magnification lightcurve.

  Microlensing can also be detected directly from the gravitational
deflection, if the positions of the images can be monitored with very
high accuracy using interferometry (H\/og, Novikov, \& Polnarev 1995;
Miyamoto \& Yoshii 1995; Gould 1996). Whereas the maximum
magnification in a microlensing event goes as $\theta^{-4}$ when the
impact parameter $\theta$ is high, the deflection decreases only as
$\theta^{-1}$. Thus, with good astrometric accuracy the effect can be
observed for very large impact parameters, increasing enormously the
probability of detecting an event.

  The very high astrometric accuracy required for microlensing can be
achieved using interferometry in the near infrared to measure the
relative position of a bright guide star (used to correct for the phase
shift caused by seeing), and a reference star located within the
isoplanatic angle of the guide star, which has a radius of $\sim 30''$,
using the technique of closure phase (??? Shao \& Colavita 1992a).
The astrometric accuracy of ground-based interferometers is at present
$\sim 50 \muas$ in the Palomar Testbed Interferometer, with guide stars
of magnitude $K \lta 6$ and reference stars with $K \lta 14$. However,
using the two Keck telescopes or the VLT, this may be improved to 10
$\muas$ and down to guide and reference stars 2 to 4 magnitudes fainter
(Shao \& Colavita 1992a,b; Shao 1996, priv. communication). In this
paper, the terms guide and reference star shall refer to any pair of
stars that can be used for relative astrometry with this technique.
It turns out that the observations required to find background stars
and search for gravitational deflection are exactly the same as what is
needed to discover planets or brown dwarfs around nearby stars. Any
program to discover planets by direct imaging near stars bright enough
to be used as guide stars will also identify any background stars
adequate as references. Monitoring the relative position of the guide
and reference stars may be done for the main purpose of discovering
planets from the angular motion they induce on the star around which
they orbit (or for measuring parallax, proper motion, or the orbit of a
binary system in either the guide or the reference star), but will also
reveal the presence of gravitational deflection.

  This paper will describe three possible applications of an astrometric
monitoring program related to microlensing: measuring the mass of the
guide star, measuring the mass of a planet or brown dwarf near the guide
star, and detecting microlensing events by other objects along the line
of sight.

\sect{2. MICROLENSING WITH ASTROMETRY}

  We first consider the probability to observe a microlensing event when
monitoring the position of a background star near another bright star.
Astrometric monitoring of such a pair of stars would usually be done for
the primary purpose of searching for planets, so the guide star will be
chosen to be nearby to maximize the angular motion caused by a planet,
and the fainter reference star will typically be more distant.

  The microlensing optical depth towards the reference star, assumed to
be at a distance $D_s$, is the fraction of the sky filled by the
Einstein radii of all the lenses along the line of sight, which we
assume to have a constant density $n = \int n(M)\, dM$, where $M$ is
the mass of the lens. The Einstein radius is
$$ \theta_E = \left( {4GM\over c^2\, D}{D_s - D\over D_s} \right)^{1/2}
~, \eqno(\new) $$
where $D$ is the distance to the lens, and the optical depth due to
lenses of mass M is
$$ \tau(M)\, dM = \int_0^{D_s} dD\, D^2\, n(M)\, \pi \theta_E^2 \, dM
= {2\pi G D_s^2 \rho(M) \over 3c^2}\, dM ~, \eqno(\new) $$
where $\rho(M)=Mn(M)$ is the density of lenses of mass $M$. This
optical depth is the probability that the reference star is within an
Einstein ring at any given time, where the total magnification is
larger than $(9/5)^{1/2}$ (Paczy\'nski 1986).

  When we are searching for microlensing with astrometric measurements,
where an event is detected if the maximum deflection is larger than
$\theta_{min}$, the maximum impact parameter that allows a detection is
$\theta = \theta_E^2/\theta_{min}$. The relevant quantity is then the
fraction of the sky where the deflection is larger than $\theta_{min}$,
which we call the ``deflection optical depth'', $\tau_d$:
$$\tau_d(M) \equiv \int_0^{D_s} dD\, D^2\, n(M)\, \pi\theta^2 =
5 (\theta_{Es}/\theta_{min})^2\, \tau(M) ~, \eqno(\new) $$
where $\theta_{Es}^2 \equiv 4GM/(c^2 D_s)$.

  As an example, we consider compact objects accounting for the dark
matter in the halo, with a local density $\rho_{h0} = 0.01
\msun\pc^{-3}$. The usual optical depth is $\tau = 2.5\times 10^{-8}
(D_s/5\kpc)^2$, and from deflection the optical depth is $\tau_d =2.3
\times 10^{-4} (M/\msun)\, (D_s/5\kpc)\, (30\muas/\theta_{min})^2$.
For known stars and dark matter in the disk, with local density
$\rho_{d0} = 0.05 \msun\pc^{-3}$ (see Gould, Flynn, \& Bahcall 1996),
an optical depth 5 times larger is inferred as long as the source star
is in the plane, so that the assumption of a constant density of lenses
is approximately valid. Because these
observations would be made in the infrared, adequate sources could
probably be found to considerably larger distances in the galactic plane
than in the visual bands. Towards the bulge, the optical depth is known
to be $\tau \sim 3\times 10^{-6}$, but the Einstein radii are smaller
($\sim 0.3\mas$, consistent with the observed durations and the velocity
dispersion that the lenses and sources should have), so the deflection
optical depth for the same events is $\tau_d \sim 3\times 10^{-4}
(30\muas/\theta_{min}^2)$. The durations of these events would be
$\theta_E/\theta_{min}$ times longer than the magnification events,
or 1 to 10 years.

  The deflection angle observed during a microlensing event when the
impact parameter is much larger than the Einstein radius is
$$ \balpha = { \theta_E^2 \over \theta_0^2 + (\mu t)^2 }
\, (\btheta_0 + \bmu t) ~,  \eqno(\new) $$
where $\btheta_0$ is the impact parameter and $\bmu$ is the proper
motion. If only this deflection is measured, only
the parameters $\theta_E^2/\theta_0$ and $\mu/\theta_0$ are obtained.
To measure the Einstein radius, an independent source of information is
needed to obtain the impact parameter. When the impact parameter is
small, the magnification is also measured and this gives the impact
parameter (and the deflection angle in (\last) is then also not exact,
breaking the degeneracy), but in most cases the impact parameter will
be too large. The maximum impact parameter where the magnification is
measurable should be $\theta_0 \simeq 3 \theta_E$, corresponding to
$A_{max}=1.008$. Thus, when the lens is unknown the only quantity
independent of the impact parameter that is measured is
$\theta_E^2/\mu$, a similar situation to the microlensing events
where only the magnification is observed, and the event duration
$t_E=\theta_E/\mu$ is the measured quantity. However, the gravitational
deflection also allows us to determine the direction of the relative
proper motion between the lens and the source. The deflection trajectory
predicted by equation (\last) must be observed with some minimum degree
of sampling and accuracy to be confident that a microlensing event has
been detected, since an apparent relative angular acceleration of the
two stars could be due to several other causes. For example, the
reference star might be a binary system (notice that this would be more
difficult to distinguish from a planet orbiting the guide star).

  In order to detect several microlensing events from
the gravitational deflection, many thousands of stars would have to be
monitored astrometrically over several years, with a frequency of a few
observations per year. The total number of stars brighter than $K=5$, to
be used as guide stars, is $\sim 40000$ (similar to stars with $V<8$),
and the probability to find a reference star brighter than $K=14$ in a
field $30''$ in radius is $\sim 15\%$ (see Fig. 5 of Shao \& Colavita
1992b). Therefore, only a few thousand pairs would be available for
these magnitude limits (although the fact that {\it both} the potential
guide stars and reference stars are concentrated to the galactic plane
would increase the number of pairs available), and probably only a
fraction of these can be observed given realistic observing times. Thus,
it seems that detecting many events can only be done with more powerful
interferometers than the present ones, and a large technological
breakthrough would be required. Probably, the positions of many
background stars would have to be measured simultaneously in crowded
fields.
Nevertheless, given that these observations will be done in any case in
order to discover planets, one should keep in mind that the possibility
to detect microlensing events is not negligible.

  The optical depth is of course increased with higher astrometric
accuracy. For a fixed mass of the lens, events of smaller deflections
would also imply longer event timescales. However, if brown dwarfs with
$M\sim 10^{-2} \msun$ account for dark matter in the disk with
$\rho_{d0}= 0.05 \msun\pc^{-3}$, then events with maximum deflection of
$1\muas$ would still have timescales of $\sim 3$ years, and optical
depth $\tau_d\sim 10^{-2}$.

  Finally, we point out that the search for microlensing events using
the deflection can detect extended objects of lower surface density
than using the magnification, down to $\Sigma_{crit}\, (\theta_{min}/
\theta_E)^2$. For objects of $M=1\msun$ at distances of a few $\kpc$,
and $\theta_{min}/\theta_E\sim 10^{-2}$, this corresponds to densities
of $\sim 10^{10} \gev\cm^{-3}$.

\sect{3. MEASUREMENTS OF GUIDE STAR MASSES}

  The position of the reference star (assumed to be much more distant
than the guide star) will also be deflected by the guide star by an angle
$\alpha = \theta_E^2/\theta$. Measurement of this deflection angle yields
the mass of the guide star, since the parallax difference of the two
stars (equal to $(D_s-D)/(D\, D_s)$) is the only other quantity that
$\theta_E$ depends on, and is accurately measured by the astrometric
monitoring (and in this case, the impact parameter is obviously known).

  In order to measure the mass of the star, observations have to be done
over a period $t \simeq \theta_0/\mu$, where $\mu$ is the proper motion
\footnote{1}{For a smaller observing time, only a small fraction of the
deflection trajectory is observed, and the difference from a linear
trajectory (which is the only information on the deflection) is of order
$(\theta_E \mu t)^2/\theta_0^3$; moreover, accelerations due to orbiting
companions cannot easily be distinguished}.
In addition, the impact parameter must
be smaller than $\theta_E^2/ \theta_{min}$, where $\theta_{min}$ is the
minimum deflection angle that is measurable. The ratio
$${\theta_E^2\over \theta_{min}\, \mu t} = 2.6\, {M\over \msun}\,
{50 \kms \over v}\, {10 \yr \over t}\, {30\muas\over \theta_{min} }\,
{D_s-D\over D_s} ~, \eqno(\new) $$
where $v$ is the transverse velocity of the guide star, will most often
be greater than unity (except for high velocity, low-mass stars, which
will rarely be bright enough for being used as guide stars). This implies
that {\it if a microlensing event with a timescale $\lta 10 \yr$ is
predicted from the known positions and proper motions of two stars
adequate for relative interferometry, then the deflection will in most
cases be measurable with present interferometry techniques}.

  The difficult challenge is to find the potential guide-reference star
pairs that are sufficiently close to produce an event on a timescale
less than some specified value $t$. To estimate the number of events
that can be predicted by searching near all possible guide stars, we
define $N(\mu, F)\, d\mu$ as the number of stars in the sky having
proper motion $\mu$, with flux brighter than $F$. If an area $(\mu t)^2$
around each star is searched for potential reference stars, with average
number density $n_{ref}$, then the expected number of pairs (each of
which will be a predicted event that can yield a mass measurement) is
$N_{pair} = \int d\mu N(\mu, F) (\mu t)^2\, n_{ref}$.

  We have used the Hipparcos Input Catalogue (Turon \etal 1992) to
estimate this number of events.
We use V magnitudes, since K magnitudes are unfortunately not
available for all bright stars. This catalogue is complete down to
$V=7.3$. Figure 1 shows the sum of $\mu^2$ over all stars
brighter than the indicated magnitudes.
When multiplied by $t^2 n_{ref}$, this yields the number of
expected pairs. We see that most of the area available for predicting
events is in stars with $\mu \gta 0.5 ''/\yr$, and it increases only
slowly with magnitude. This can be simply understood as follows:
Stars with $\mu \gta 0.1 {\rm arcsec}/\yr$
(corresponding to $50\kms$ at $D=100\pc$) should be nearby disk stars,
or spheroid and thick disk stars at distances larger by no more than a
factor $\sim 5$. Thus, their density and velocity distribution can be
approximated as constant, which implies that
$N(\mu,F) d\mu = G(F/\mu^2)\, (d\mu/\mu^4) $, where
$G$ is a convolution of the cumulative luminosity function with the
distribution of transverse velocities. The expected number of events
goes as $\int (d\mu/\mu^2) G(F/\mu^2)$, which will converge at low $\mu$
when $G$ decreases faster than $\mu^{-1}$ with decreasing $\mu$, or
equivalently when the cumulative luminosity function is steeper than
$\phi(L) \sim L^{-1/2}$. This slope is achieved for luminosities $L\sim
0.1 - 1 \lsun$ in the V band (and lower luminosities in the K band),
which for a limiting magnitude $V=8$ would be at $D=20 \pc$, with
typical proper motion $0.5 ''/\yr$.

   For an observing time $t=10\yr$, the number of events that can be
predicted is $\sim N_{ref}/10^7$, where $N_{ref}$ is the total number
of stars above the magnitude limit for reference stars (notice that
high proper motion stars are isotropically distributed in the sky, so
only the total number of reference stars available is relevant here).
The total number of stars is $1.8\times 10^8$ down to $V=17$, and
$6.5\times 10^8$ down to $V=20$ (see Table 4.2 in Mihalas \& Binney
1981); the numbers are probably similar above $K=14$ and $K=17$,
respectively (see also Figure 5 in Shao \& Colavita 1992b). Thus, even
with the limit $K=14$, we would expect to
predict $\sim 20$ events leading to mass measurements. Most of the
events will be caused by nearby stars, which would be likely candidates
for astrometric monitoring in a planetary search program in any case.
  
\sect{4. MEASUREMENT OF PLANET AND BROWN DWARF MASSES}

  The major reason to conduct astrometric monitoring programs will be
to discover planets from the angular motion induced on the guide star.
This angular motion is proportional to the mass of the planet, so the
most massive planets are likely to be discovered. Giant planets and
brown dwarfs may also be discovered by direct imaging. Mass measurements
of these objects are important for testing theories of their structure
and evolution (e.g., Marley \etal 1996). If the angular motion of the
star is observed for a sufficiently long time to determine the orbit the
mass of the planet can be derived, but the orbital period may be very
long. So far, several giant planets have been
discovered from radial velocity measurements (Mayor \& Queloz 1995;
Marcy \& Butler 1996; Butler \& Marcy 1996) and a brown dwarf by imaging
(Nakajima \etal 1995, Oppenheimer \etal 1995). The brown dwarf (Gl 229B)
should have an orbital period of several hundred years.

  The mass of a companion of a guide star could also be measured from
its gravitational deflection of the light of the reference star. The
best case is for brown dwarfs. Assuming a mass $M_{bd}=0.03 \msun$ at a
distance of $10$ pc, the brown dwarf Einstein radius is $5\mas$, and at
a separation of 1'' the deflection angle is $25 \muas$. The luminosity of
this brown dwarf would be as low as $5\times 10^{-7} \lsun$ (or $K=17$
at 10 pc) unless it is younger than a few billion years (e.g., Marley
et al. 1996), so it should generally be detectable. Assuming the
brown dwarf is detected prior to the microlensing event, the time and
the maximum deflection of the event could be predicted.

  For the low masses of brown dwarfs and giant planets, the maximum
impact parameter of an event is probably limited by the astrometric
accuracy, rather than the proper motion and the observing time.
Therefore, the area where a background star would produce an observable
deflection is $2\mu t\, \theta_{\max}$, or $20$ arcsec$^2$ for typical
brown dwarf parameters, giving an average probability of $1\%$ for an
event for a magnitude limit $K=18$. The brown dwarf GL229B has $\mu =
0.74 ''/\yr$, $D=7 \pc$, and for a mass $M=0.03\msun$ and $\theta_{min}
= 10 \muas$ an area of $40$ arcsec$^2$ is swept by every ten years.
Since the galactic latitude is $15\degree$, the chance to find an
adequate reference star in this area is not very small.

  Discovering a previously unknown planet from the gravitational
deflection is much more difficult, because many stars would have to be
frequently monitored to search for rare and short events if they are
not predicted in advance. 
However, the discovery of planets with microlensing over a wide mass
range (down to much lower masses than those under discussion here) in
distant systems using the traditional technique of measuring
magnification lightcurves is very promising, as has been thoroughly
discussed (Mao \& Paczy\'nski 1991; Gould \& Loeb 1992; Bennett \& Rhie
1996).

\sect{5. CONCLUSIONS}

  Accurate astrometric monitoring of pairs of guide and reference stars
with interferometry will determine proper motions and parallaxes of high
precision, and reveal extrasolar planets. Exactly the same observations
should reveal gravitational deflection of the reference star (which is
normally much more distant than the guide star) by the guide star, by
any orbiting companions of the guide star, and by other objects near the
line of sight to the reference star.

  Initially, most of the stars that will be monitored will be nearby,
because the angular motion caused by planets is larger for nearby stars.
These are also the more likely stars to have adequate reference stars
allowing for a mass measurement, given their high proper motions. They
are also the best candidates for measuring the mass of a companion
brown dwarf, for the same reason (notice also that the deflection angle
at a fixed angular impact parameter scales as the inverse of the
distance). To detect microlensing by other objects,
the most important consideration is to find a distant reference star,
which increases the optical depth. Any guide star is equally good (in
fact, distant ones are best because events involving the guide star
could also be detected). Although distant guide stars would not be
chosen for discovering planets, their astrometric monitoring is also
interesting for high precision measurement of parallaxes and proper
motions.

  At present, the detection of gravitational deflection is still
difficult, and probably only a handful of stellar masses may be
determined in the next $\sim 10$ years. Of course, the number of events
that can be detected increases enormously with the astrometric accuracy
and the total length of time of observation. Many binary star orbits are
only known to us today because of the observations done over periods of
100 years or longer. Over the long term, the use of gravitational
deflection to measure masses is likely to become of fundamental
importance in astronomy.

\bigskip
{\bf Acknowledgements}
I thank Andy Gould for many comments that resulted in a much improved
paper, and John Bahcall, Shri Kulkarni, Bohdan Paczy\'nski and
Michael Shao for useful comments.
This work has used the Hipparcos Input Catalog, which was extracted from
the ESO/ST-ECF/CADC astronomical database.
This research was supported by a W. M. Keck Foundation grant.

\vskip 1cm

\vfill\eject

\parskip .0cm plus .03cm

\sect{REFERENCES}

\refbook{Alard, C. 1996, in {\it Proc. IAU Symp. 173}, p. 215, eds. C. S.
Kochanek, J. N. Hewitt; Kluwer Academic Publishers,
Dordrecht/Boston/London}
\physrevl{Alcock, C., et al. 1995}{74}{2867}
\refbook{Alcock, C., et al. 1996a, ApJ, in press (astro-ph 9512146)}
\refbook{Alcock, C., et al. 1996b, submitted to ApJ}
\refbook{Bennett, D. P., \& Rhie, S. H. 1996, submitted to ApJ
(astro-ph 9603158)}
\refbook{Butler, R. P., \& Marcy, G. W. 1996, ApJ. Let., in press}
\apj{Gould, A., \& Loeb, A. 1992}{396}{104}
\refbook{Gould, A. 1996, PASP, 108, in press (astro-ph 9604014)}
\apj{Gould, A., Flynn, C., \& Bahcall, J. N. 1996}{465}{in press}
\apj{Griest, K. 1991}{366}{412}
\apj{Han, \& Gould, A. 1996}{468}{in press}
\aa{H\/og, E., Novikov, I. D., \& Polnarev, A. G. 1995}{294}{287}
\apj{Kiraga, M. \& Paczy\'nski, B. 1994}{430}{101}
\apj{Mao, S., \& Paczy\'nski, B. 1991}{388}{L45}
\refbook{Marcy, G. W., \& Butler, R. P. 1996, ApJ. Let., in press}
\refbook{Marley, M. S., Saumon, D., Guillot, T., Freedman, R. S.,
Burrows, A., \& Lunine, J. I. 1996, submitted to Nature}
\nature{Mayor, M., \& Queloz, D. 1995}{378}{355}
\refbook{Mihalas, D., \& Binney, J. 1981, {\it Galactic Astronomy}
(W. H. Freeman and Co.: San Francisco)}
\aj{Miyamoto, M., \& Yoshii, Y. 1995}{110}{1427}
\nature{Nakajima, T., et al. 1995}{378}{463}
\science{Oppenheimer, B. R., Kulkarni, S. R., Matthews, K., \& Nakajima, T.
1995}{270}{1478}
\apj{Paczy\'nski, B. 1986}{304}{1}
\apj{\refrule ~ 1991}{371}{L63}
\actastr{\refrule ~ 1995}{45}{345}
\refbook{\refrule ~ 1996, ARA\&A, in press (astro-ph 9604011)}
\araa{Shao, M., \& Colavita, M. M. 1992a}{30}{457}
\aa{\refrule ~ 1992b}{262}{353}
\aa{Turon, C., G\'omez, A., Crifo, F., Creze, F., Perryman, M. A. C.,
Morin, D., Arenou, F., Nicolet, B., Chareton, M., \& Egret, D. 1992}{258}{74}
\apj{Udalski, A., et al. 1994a}{436}{L103}
\apj{Udalski, A., Szyma\'nski, M., Kaluzny, J., Kubiak, M., Mateo, M.,
Krzemi\'nski, W. 1994b}{426}{L69}
\actastr{Udalski, A., et al. 1994c}{44}{227}
\apj{Zhao, H. S., Spergel, D. N., \& Rich, R. M. 1995}{440}{L13}

\vfill\eject

\topinsert
\centerline{
\epsfxsize=5.0in \epsfbox{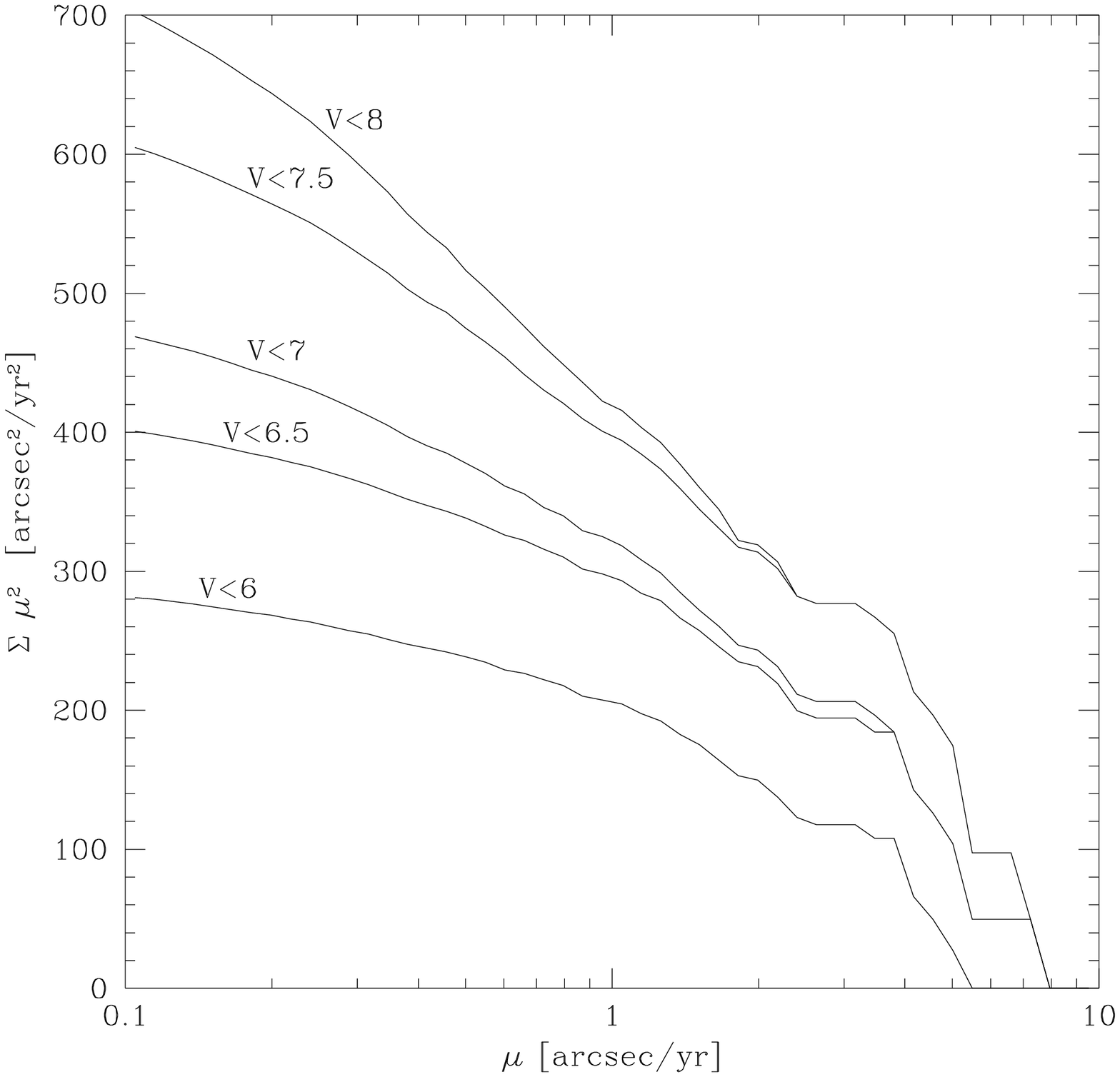}
}

\baselineskip=12truept \leftskip=3truepc \rightskip=3truepc 
\noindent {\bf Figure 1:}
The vertical axis shows the sum of the square of the
proper motions over all stars in the Hipparcos Input Catalog
brighter than V=(8,7.5,7,6.5,6), and with proper motion higher than
$\mu$. When multiplied by the square of the observing
time in years, and by the average density of reference stars in
arcsec$^{-2}$, this yields the expected number of stars whose mass can
be measured from the gravitational deflection of the reference star
light.
\endinsert

\vfill
\eject
\bye